\documentclass{aa}  
\usepackage{graphicx}
\usepackage{txfonts}
\usepackage{longtable}
\begin{document}
   \title{A new young stellar cluster embedded in a molecular cloud in the far outer Galaxy\thanks{Based on observations collected at the ESO 8.2-m VLT-UT1 Antu telescope (program 65.C-5740).}}


   \author{Jo\~ao L. Yun\inst{1}
		\and
		Ana Lopez-Sepulcre\inst{2}
          \and
          Jos\'e M. Torrelles\inst{3}
          }

   \offprints{Joao L. Yun}

   \institute{Universidade de Lisboa - Faculdade de Ci\^encias \\
Centro de Astronomia e Astrof\'{\i}sica da Universidade de Lisboa, \\
Observat\'orio Astron\'omico de Lisboa, \\
Tapada da Ajuda, 1349-018 Lisboa, Portugal\\
             \email{yun@oal.ul.pt}
	\and
Departament d'Astronomia i Meteorologia, \\
Facultat de F\'{\i}sica, Planta 7a, Universitat de Barcelona,\\ Avenida Diagonal 647, 08028 Barcelona, Spain \\
             \email{alopezse13@fis.ub.edu}
         \and
             Instituto de Ciencias del Espacio (CSIC) and Institut d'Estudis Espacials de Catalunya, \\ Facultat de F\'{\i}sica, Planta 7a, Universitat de Barcelona,\\ Avenida Diagonal 647, 08028 Barcelona, Spain \\
             \email{torrelles@ieec.fcr.es}
             }

   \date{Received March 29, 2007; accepted }

 
  \abstract
   {}
   {We report the discovery of a new young stellar cluster and molecular cloud located in the far outer Galaxy, seen towards IRAS~06361-0142, and we characterise their properties.}
   {Near-infrared images obtained with VLT/ISAAC through $JHK\!s$ filters,  millimetre line observations of CO(1-0) obtained with SEST, and VLA 6 cm continuum maps obtained from archive data.}
   {The cloud and cluster are located at a distance of 7 kpc and a 
Galactocentric distance of 15 kpc, well in the far outer Galaxy. Morphologically, IRAS~06361-0142 appears as a cluster of several tens of stars surrounded by a nearly spherical nebular cavity centred at the position of the IRAS source.  The cluster appears composed of low and intermediate-mass, young reddened stars with a large fraction having cleared the inner regions of their circumstellar discs responsible for $(H-K\!s)$ colour excess. The observations are compatible with a 4 Myr cluster with variable spatial extinction between $A_v=6$ and $A_v=13$. 
}
   {}


\titlerunning{ A new young stellar cluster in the far outer Galaxy  }
\authorrunning{Yun, Lopez-Sepulcre \& Torrelles}

   \maketitle
%

\section{Introduction}

Stars form in molecular clouds across the Galaxy. Most molecular gas and dust exist in the Galactic disc and are most abundant 
toward the inner Galaxy (\cite{clemens88}).
Thus, most star formation sites have been found and studied in regions of the inner disk and in nearby molecular clouds (e.g. \cite{tapia91}; \cite{strom93}; \cite{mccaughrean94}; \cite{horner97}; \cite{luhman98}).
A much smaller number of star formation sites are known toward the outer Galaxy and at large Galactocentric distances ($>$13.5 kpc). 
Discovery of isolated young stellar objects located in the outer Galaxy at an estimated Galactocentric distance of about 15-19 kpc has been reported by \cite{kobayashi00}. \cite{santos00} reported the discovery of the two most distant  Galactic young stellar clusters (located at a distance of 10.2 kpc and a Galactocentric distance of 16.5 kpc) embedded in a molecular cloud containing a CS dense core.  \cite{snell02} presented a study of the molecular cloud content and star formation activity in the far outer Galaxy, which they define as the region with Galactocentric distances greater than 13.5 kpc, a definition based on the radial distribution of CO emission (\cite{digel96}, \cite{heyer98}). They identify 11 stellar clusters in the second Galactic quadrant with Galactocentric distances between 13.5 and 17.3 kpc. 

However, there are still relatively few known resolved young embedded stellar clusters in the far outer Galaxy. Also, the small number of known star formation sites in the far outer Galaxy has not allowed a systematic analysis of the star formation properties in regions very far from the Galactic centre. These rather different environments (lower metallicity, lower gas density and ambient pressure) could result in different rates of cloud evolution affecting the star formation process. This in turn may result in a different distribution of stellar masses in a cluster (different IMF) and consequently a different feedback mechanism from the stellar to the molecular component.

We have begun a study of young embedded clusters in the third Galactic quadrant and seek to characterise cloud properties and star formation activity as a function of Galactocentric distance in the outer Galaxy. As part of this study, we report here the discovery of a new young stellar cluster embedded in a molecular cloud located far in the outer Galaxy, seen towards IRAS~06361-0142.
At a distance of about 7 kpc (see below), it is one of the most distant young clusters known to date.

Section~2 describes the observations and data reduction. In Sect.~3, we present and discuss the results. A summary is given in Sect.~4.


\section{Observations and data reduction}

\subsection{Infrared observations}

Near-infrared ($J$, $H$ and $K\!s$) observations were conducted during a night of excellent seeing (0.35 arcsec in the $K\!s$-band), 2000 November 10, using the ESO Antu (VLT Unit 1) telescope equipped with the short-wavelength arm (Hawaii Rockwell) of the ISAAC instrument. The ISAAC camera (\cite{moorwood98}) contains a 1024 $\times$ 1024 pixel near-infrared array and was used at a plate scale of 0.147 arcsec/pixel resulting in a field of view of 2.5~$\times$~2.5 arcmin$^2$ on the sky.  At each of 6 different jitter positions (with a jitter box of 30 arcsec), series of 15 images with 4~s exposure time were taken. 

Standard procedures for near-infrared image reduction were applied (e.g. \cite{yun94}) resulting in a final mosaic image for each band. A sky image was computed for each dithered frame. Each frame was then sky subtracted and flat-fielded. The final clean images were corrected for bad pixels while constructing the final mosaic, which covers about $3.1 \times 3.1$ arcmin$^2$ on the sky. Due to the presence of significant field distortion in these ISAAC images (two pixels at the edges and 2.5 pixels in the corners), the individual tiles of mosaics were corrected before they were tiled together. The correction for field distortion was performed using IRAF/GEOTRAN, together with the adequate correction files provided by ESO at their web page.

Source extraction and aperture photometry were performed on the final mosaic image using IRAF packages. The final calibration was performed by choosing bright isolated stars on the image and transforming to the 2MASS photometric system. 
The numbers of stars detected in each band were 557, 644, and 785, in the $J$, $H$, and $K\!s$ bands, respectively.
The completeness limits of our images is estimated to be at magnitudes $J=20.0$, $H=19.5$, $K\!s=19.0$. 
Photometry errors range from about 0.08 mags for the bright stars to 0.15 mags for the fainter ones.

\subsection{ CO(1-0) line observations}

As part of a CO survey of IRAS-based protostars, millimetre observations at 115 GHz (the J=1-0 rotational transition of CO) were carried out toward IRAS~06361-0142 using the 15m Swedish-ESO Submillimetre Telescope (SEST) in Chile during September 1999. The source was observed using dual beam switching (chop throw of 11 arcmin) with integration times of 1 minute. 
The half-power beamwidth of the telescope and the main-beam efficiency were 45 arcsec and 0.65, respectively.  We used SIS receivers and the SEST
acousto-optical spectrometer as a back-end, with a resolution of about 0.11~km~s$^{-1}$ at 115 GHz. The pointing error was found to be better than 6 arcsec. 

Spectral line intensities were calibrated and corrected for atmospheric losses using the standard chopper wheel method to obtain the antenna 
temperature $T^{\ast}_A$.
Baselines were fitted and removed using standard procedures of the Continuum and Line Analysis Single-dish Software (CLASS) package developed at the
Observatoire de Grenoble and IRAM Institute.

\subsection{VLA centimetre observations}

We searched for continuum observations made with the Very Large Array (VLA) of the National Radio Astronomy Observatory (NRAO)\footnote{The National Radio Astronomy Observatory is a facility of the National Science Foundation operated under cooperative agreement by Associated Universities, Inc.} 
to look for a radio continuum counterpart to IRAS~06361-0142.
The VLA data archive contained 6 cm continuum observations towards IRAS 06361-0142, which were carried out with the VLA in its B configuration on 2005 May 8. The observations were made in both left and right circular polarization, with an effective bandwidth of 100 MHz. The phase centre of the observations was $\alpha(2000)$ = 6h38m39.4s, $\delta (2000) =-1^{\circ} 44' 49''\!\!.0$. The data were reduced and calibrated using the Astronomical Image Processing System (AIPS) software of the NRAO. The flux calibrator was 3C138, with an adopted flux density of 3.7 Jy, while the phase calibrator was 0725-009, with a bootstrapped flux density of 1.11$\pm$0.01 Jy. The image had an rms sensitivity of 140 $\mu$Jy/beam, with a size (FWHM) and position angle (P.A.) of the synthesized beam of $1''\!\!.8\times 1''\!\!.4$ and $-20^{\circ}$, respectively.

\section{Results and discussion}

\subsection{Molecular gas and distance}

   \begin{figure}
   \centering
   \includegraphics[width=8cm]{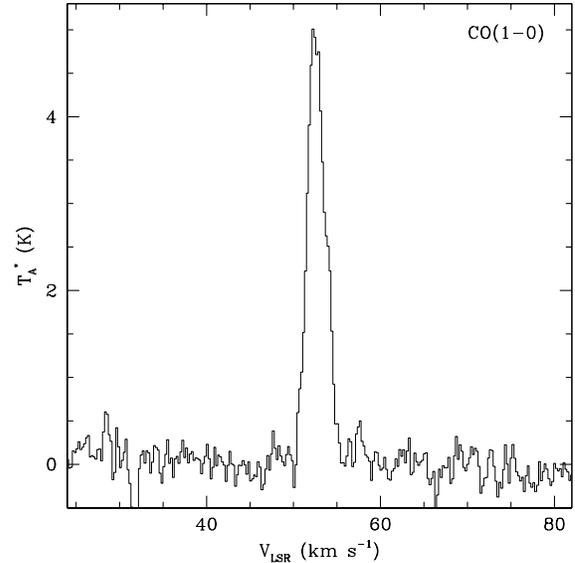}
   \caption{
CO (J=1-0) spectrum towards IRAS~06361-0142 displaying a line at an LSR velocity of about 53~km~s$^{-1}$. The spectrum has been smoothed to a resolution of 0.22~km~s$^{-1}$.
	} 
	\label{fig1}%
    \end{figure}
%

In Fig.~1, we present the SEST CO (J=1-0) central spectrum obtained towards IRAS~06361-0142. A CO line is detected at an LSR velocity of about 53 km s$^{-1}$. 
Assuming a Galactic rotation curve that is essentially flat beyond the solar circle (\cite{clemens85}), and a Galactocentric distance to the Sun ($R_o$) of 8.5 kpc, we derive a kinematic distance of 7 kpc to this molecular cloud. 
Given the Galactic longitude, this implies a Galactocentric distance of 15.0 kpc to the cloud, well into the far outer Galaxy and among the largest within known Galactic star formation sites. We estimate a distance uncertainty of about 20\% due to uncertainties in the rotation curve and possible streaming motions (e.g. for values of 10--15 km s$^{-1}$ for streaming motions, either red or blueshifted relative to the circular velocity, we get distances of 5.3--5.9 or 8.5--9.2 kpc, respectively).

\subsection{Extended nebula}

A careful analysis of the Digitized Sky Survey (DSS - red plate) at the position of IRAS~06361-0142 reveals a fuzzy spot at this location. The fuzziness could represent extended, diffuse nebulosity, or stellar sources with confusion due to insufficient resolution and low sensitivity of the DSS.
2MASS images look similar showing a relatively bright, diffuse stellar-like source.

   \begin{figure}
   \centering
   \includegraphics[width=9cm]{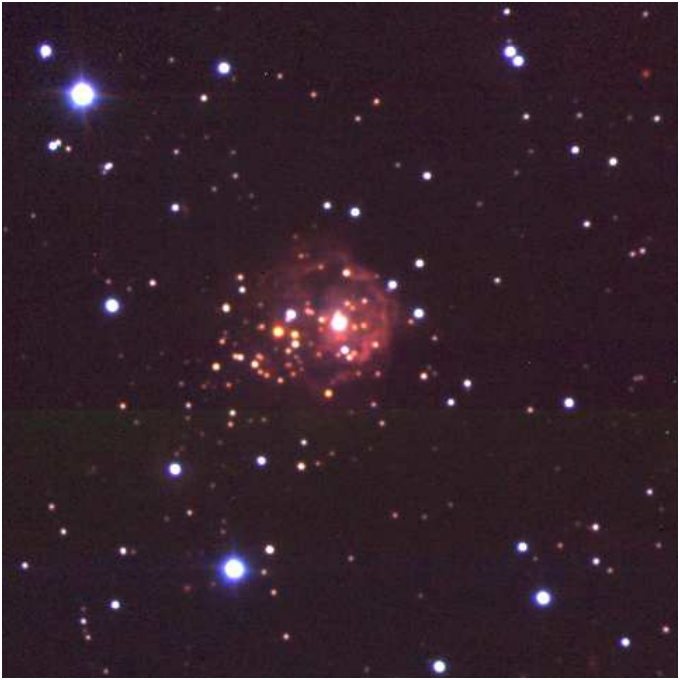}
   \caption{
J (blue), H (green), and Ks (red) colour composite image towards IRAS~06361-0142. The image covers about $70\times 70$ arcsec$^2$. North is up and East to the left. 
	} 
	\label{fig2}%
    \end{figure}

Figure~2 presents the $JHK\!s$ colour composite image (on-line) obtained towards IRAS~06361-0142. A group of stars is seen surrounded by extended nebulosity with a nearly spherical shape. The excellent seeing at the time of the observations is evident in this image: despite the small angular size, the colour image clearly conveys the impression of depth and the nebular region is seen as being composed by patchy clouds. Morphologically, it bears a striking similarity to other star formation nebulae that display a cluster of stars near the centre of a nearly spherical cavity defined by illuminated walls of a surrounding molecular cloud. Examples are the Rosette nebula (e.g. \cite{ybarra04}) and N90 (in the SMC, \cite{nota07}). However, unlike these examples, no HII region has been detected with the VLA (Sect. 2.3). Thus, if no ionization ocurred, the origin of such a morphological structure, with the partial clearing of the central cavity, may have happened through the action of stellar winds from young stars. The nebula seen is likely to be reflected light off the walls of the nearly spherical cavity.

   \begin{figure}
   \centering
   \includegraphics[width=8cm]{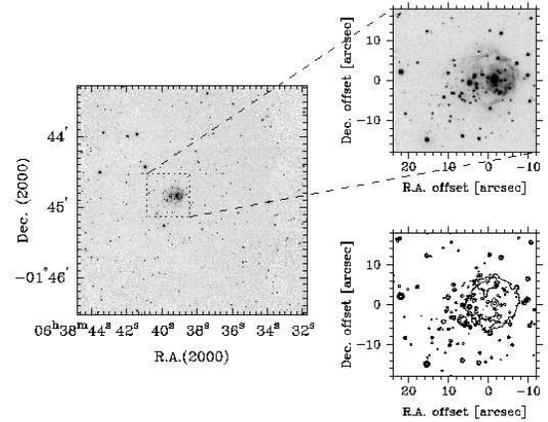}
   \caption{
$K\!s$-band image towards IRAS~06361-0142. The close-up view at the top shows the central $36'' \times 36''$ region (the ``cluster region'').
The close-up panel at the bottom shows the isophotes of the same region where the very close to circular shape of the extended nebulosity is evident.  The axes give the coordinates relative to the estimated centre of the nebulosity, nearly coincident with the IRAS source (see text). 
The contour levels are at 5$\sigma$, 25$\sigma$, and 250$\sigma$ 
above the sky level, where $\sigma$ is the sky background noise in the central co-added region. 
	} 
	\label{fig3}%
    \end{figure}
%

The Ks-band image is shown in Figure~3 with the top and bottom right panels containing zoomed-in (close-up) views. The shape of the isophote of the nebulosity seen at the bottom right panel is very close to a circle. We estimate the position of the centre of this circle to be located about 3$''$ east of the IRAS source, well-contained in the IRAS ellipse error. In fact, given the large size ($26''$) of the IRAS error ellipse in the R.A. direction, we assume a coincidence of the centre of the nebula and the position of the IRAS source.  

The angular diameter of the nebula (measured to about 3 sigma above the sky background) is 20 arcsec (0.7 pc at 7 kpc).

\subsection{Reddening}

   \begin{figure}
   \centering
   \includegraphics[width=8cm]{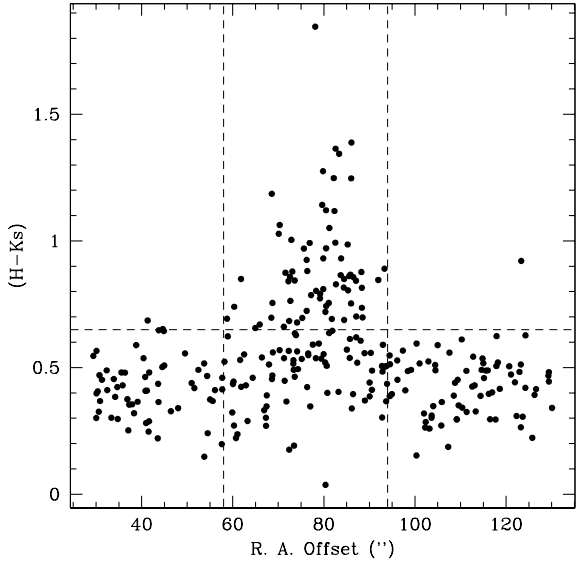}
   \caption{
Plot of the observed $(H-K\!s)$ colour of stellar sources as a function of the R.~A. offset relative to the lower left (SE corner) of Fig.~3 (main panel). Sources closer than about 20 arcsec to the borders were excluded. The reddening effect ($(H-K\!s\geq 0.65$, horizontal dashed line) of a dense cloud core is clearly seen for values of R.~A. offsets between approximately 58$^{\prime\prime}$ and 94$^{\prime\prime}$ (indicated by vertical dashed lines).
	} 
	\label{fig4}%
    \end{figure}
%

Using the results of the point source photometry of the near-IR images, we present (in Fig.~4) the $(H-K\!s)$ colour of the stellar sources as a function of the R.~A. offset relative to the lower left (SE corner) of Fig.~3 (a similar plot is obtained as a function of declination offset). 
Along the horizontal axis, for all values of R.~A., there are values of $(H-K\!s)$ between about 0.2 and about 0.65. The reddening effect of the cloud clearly stands out as enhanced values of $(H-K\!s)$ for values of R.~A. offset between approximately 58$^{\prime\prime}$ and 94$^{\prime\prime}$ (a range of about 36 arcsec -- about 1.2 pc at the source distance of 7 kpc). 
For the general case, this colour excess of $(H-K\!s)=0.65$ would just separate the objects into two groups: the group of foreground stars and the group of \{embedded + background\} stars. However, given the fact that at this distance and location in the Galaxy, most visible stars are either foreground or embedded in the cloud (with very few being background stars), this reddening effect at this region of the sky separates foreground from embedded stars and thus effectively selects cluster members.

Consequently, we take this diagram of Fig.~4 and the association of CO emission 
as evidence of a cloud core and use it to define the ``cluster region'' as this central region of about 36 arcsec on each side. The cluster region is shown in the close-up panels of Fig.~3. A star that is located in the cluster region {\it and\/} has $(H-K\!s)>0.65$ will be considered here to be a cluster member. Evidently, that is a sufficient but not a necessary condition. There may be additional cluster members outside the cluster region or with bluer values of $(H-K\!s)$. Thirty-four sources fit these criteria and are thus good candidate cluster members. Photometry of these sources is given in Table~1\footnote{Table 1 is available at the Centre de Donn\'ees Astronomiques de Strasbourg (CDS), at http://cdsweb.u-strasbg.fr.}.

\begin{table}
\caption{Photometry of candidate cluster sources}
\label{table:1}      
\centering                          
\begin{tabular}{c c c c c }        
\hline\hline                 
R.A. & Dec & m$_{K\!s}$ & $(H-K\!s)$ & $(J-K\!s)$ \\
(2000) & (2000) &  & &  \\
\hline                        
\noalign{\vspace{2 pt}}
06 38 39.38 &-$\,$01 45  04.1 &  16.7 &  0.97 &  2.27 \\
06 38 39.88 &-$\,$01 44  59.7 &  18.1 &  0.88 &  2.33 \\
06 38 40.63 &-$\,$01 44  57.8 &  17.8 &  0.85 &  2.12 \\
06 38 39.19 &-$\,$01 44  56.5 &  16.5 &  1.34 &  3.29 \\
06 38 39.53 &-$\,$01 44  55.1 &  18.0 &  0.80 &  2.31 \\
06 38 38.52 &-$\,$01 44  54.7 &  17.1 &  0.89 &  2.21 \\
06 38 39.62 &-$\,$01 44  54.6 &  18.0 &  0.99 &  2.78 \\
06 38 38.84 &-$\,$01 44  54.4 &  18.5 &  0.70 &  1.85 \\
06 38 38.98 &-$\,$01 44  54.4 &  18.8 &  0.86 &  2.37 \\
06 38 39.67 &-$\,$01 44  54.2 &  18.1 &  0.93 &  2.44 \\
06 38 39.98 &-$\,$01 44  53.7 &  16.8 &  0.87 &  2.46 \\
06 38 39.47 &-$\,$01 44  53.5 &  18.8 &  0.79 &  2.21 \\
06 38 39.71 &-$\,$01 44  53.4 &  16.6 &  0.97 &  2.68 \\
06 38 39.81 &-$\,$01 44  52.7 &  17.2 &  0.68 &  2.00 \\
06 38 39.04 &-$\,$01 44  52.7 &  17.3 &  0.86 &  2.14 \\
06 38 39.12 &-$\,$01 44  52.4 &  18.5 &  0.82 &  1.61 \\
06 38 39.47 &-$\,$01 44  52.0 &  17.8 &  0.77 &  2.32 \\
06 38 38.86 &-$\,$01 44  51.4 &  17.5 &  0.88 &  2.32 \\
06 38 39.42 &-$\,$01 44  51.3 &  16.6 &  0.81 &  2.45 \\
06 38 39.93 &-$\,$01 44  51.0 &  18.1 &  0.84 &  2.18 \\
06 38 39.42 &-$\,$01 44  50.3 &  16.3 &  1.27 &  3.56 \\
06 38 39.54 &-$\,$01 44  50.0 &  14.5 &  1.85 &  4.46 \\
06 38 39.66 &-$\,$01 44  49.5 &  18.1 &  0.88 &  2.69 \\
06 38 38.94 &-$\,$01 44  49.4 &  18.1 &  0.84 &  2.63 \\
06 38 39.12 &-$\,$01 44  49.1 &  13.8 &  0.85 &  2.31 \\
06 38 38.61 &-$\,$01 44  49.0 &  18.3 &  0.85 &  2.09 \\
06 38 39.23 &-$\,$01 44  48.7 &  17.5 &  0.83 &  2.16 \\
06 38 39.91 &-$\,$01 44  47.6 &  16.7 &  0.76 &  2.19 \\
06 38 39.06 &-$\,$01 44  47.1 &  17.9 &  0.81 &  2.12 \\
06 38 40.42 &-$\,$01 44  44.9 &  17.5 &  0.66 &  1.74 \\
06 38 40.73 &-$\,$01 44  44.8 &  18.8 &  0.74 &  1.82 \\
06 38 38.85 &-$\,$01 44  44.2 &  19.2 &  0.81 &  1.96 \\
06 38 39.06 &-$\,$01 44  43.8 &  16.7 &  0.99 &  2.55 \\
06 38 39.42 &-$\,$01 44  40.0 &  19.4 &  0.93 &  2.51 \\
\hline\hline                                
\end{tabular}
\end{table}


   \begin{figure}
   \centering
   \includegraphics[width=8cm]{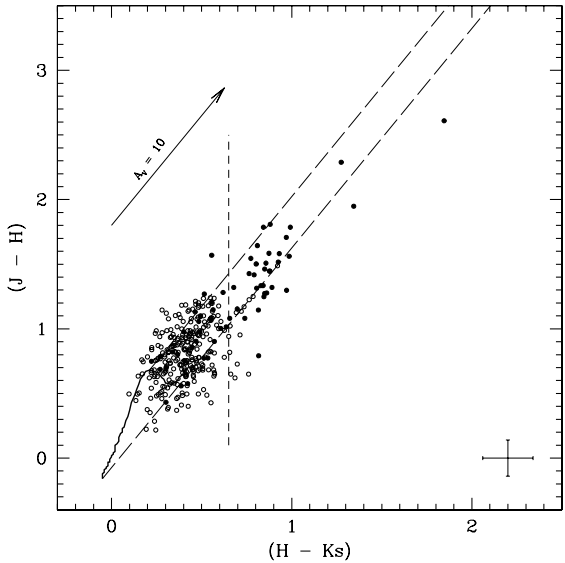}
   \caption{
Near-infrared colour-colour diagram of the region towards IRAS~06361-0142 seen in Fig.~3. 
Filled circles represent the 34 sources contained in the $36''\times 36''$ central region (``cluster region'', where most sources are likely to be members of the cluster -- see text). 
Open circles are sources outside the cluster region.
The solid line represents the locii of unreddened main-sequence stars, while long-dash lines indicate the reddening band. The reddening vector indicates the direction of the shift produced by extinction by dust with standard properties. 
The location of the vertical dotted line, derived from Fig.~4, is at $(H-K\!s) = 0.65$.  
The cross in the lower right corner represents the average error bars.
	} 
	\label{fig5}%
    \end{figure}
%

\subsection{Young stellar objects}

In the region shown in the main panel of Fig.~3, we have detected a total of 557, 644, and 785 point sources, respectively in the $J$, the $H$, and the $K\!s$-bands. Using the point sources detected simultaneously in all three bands, 
we have plotted the near-infrared colour-colour diagram, $(J-H)$ versus $(H-K\!s)$, shown in Figure~5.
In this plot, the location of the vertical dotted line, derived from Fig.~4, is at $(H-K\!s) = 0.65$. Most of the sources to the right of this line ($(H-K\!s) > 0.65$) are  in the cluster region.
Sources to the left of the dotted line are both inside and outside the cluster region.

Most stars are located within the reddening band where stars appear if they are main-sequence stars reddened according to the interstellar extinction law(\cite{rieke85}), which defines the reddening vector (traced here for $A_V=10$).
Pre-main-sequence YSOs that have had time to clear the inner regions of their circumstellar discs lie in this region as well. Giant stars appear slightly above this band. Stars located to the right of the reddening band are likely to be embedded young star objects with infrared excess emission due to the presence of circumstellar material (\cite{adams87}). 
This is the case of the reddest object (in both colour indices) easily recognised in Fig.~2 by its very red colour.

Notice that few sources lie in the infrared excess region. This lack of sources in this region suggests that most sources ($> 50$\%) have dissipated the inner parts of their circumstellar discs that are responsible for the $(H-K\!s)$ excess. This result indicates that this cluster is not extremely young, setting a lower limit of 3 Myr for its age. This is because inner-disc frequencies in young clusters have been found to steadily decrease as the cluster ages approach about 6 Myr (\cite{haisch01}), with about 50\% of the sources exhibiting excess emission from circumstellar discs at 3 Myr and very few sources at ages greater than about 6 Myr.

   \begin{figure}
   \centering
   \includegraphics[width=8cm]{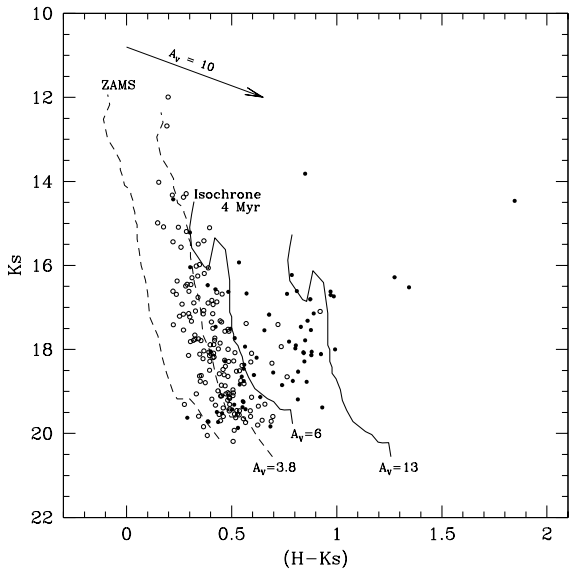}
   \caption{
Colour-magnitude diagram of the region towards IRAS~06361-0142 seen in Fig.~3.
Filled circles and open circles have the same meaning as in Fig.~5. The two dashed lines are, respectively, the zero-age main-sequence (ZAMS) for an arbitrary distance with zero extinction and with $A_v=3.8$. The two solid lines are the 4 Myr isochrone with extinctions $A_v=6$ and $A_v=13$. The reddening vector indicates the direction of the shift produced by extinction by dust with standard properties.
 	} 
	\label{fig6}%
    \end{figure}
%

An additional estimate of the age (or range of ages) for this cluster can be obtained from the colour-magnitude diagram ($K\!s$, $(H-K\!s)$) plotted in Fig.~6. In this plot, the extincted ZAMS represents a fit to the open circles (which are spread mostly along a nearly vertical band) including the two brightest sources which are clearly blue, field sources (non-cluster members).
This value of extinction represents a lower limit to the extinction in the region and is is due to the foreground diffuse ISM.

Focusing our attention now on the filled circles (assumed to be cluster members), the location of a source on this colour-magnitude diagram is the result of a combination of cluster distance, cluster age, and extinction of the line-of-sight towards the source. 
Assuming a fixed distance for the cluster members, vertical spread (scatter in magnitude $K\!s$) can be caused mostly by a spread in age. Variable extinction will only produce small vertical dispersion because the extinction vector is almost horizontal (has a low slope, see Fig.~6).
On the other hand, horizontal spread in this diagram (scatter in colour)  can be due to different ages, to variable spatial extinction, or to a combination of both. 
Analysis of Fig.~2 strongly suggests the existence of variable extinction, with highly different amounts of dust across the cluster. Non-coeval star formation cannot be excluded either. Thus, we have tried to fit the filled circles using pre-main-sequence isochrones reddened by variable amounts of extinction. 

Fitting pre-main-sequence isochrones to young stellar clusters is a rather uncertain process but it can give a rough estimate of the age or a range of ages of a young cluster.  We have used the kinematic distance of 7 kpc and the set of \cite{siess00} isochrones with masses ranging from 0.1 to 7 M$_{\odot}$. 
Given the absence of metallicity information, we adopted solar metallicity. A set of isochrones of different ages and extinctions were generated and plotted on the colour-magnitude diagram. Figure~6 shows a possible fit to the filled circles: an isochrone of 4 Myr with extinctions varying from $A_v=6$ to $A_v=13$. 
Isochrones of younger ages will fit only the lower mass-end (the fainter sources). They would require the absence of any intermediate-mass stars restricting the cluster population to only low-mass stars. In fact, assuming an age of 4 Myr for this cluster, and considering the faintest cluster member, the lowest mass detected in our $K\!s$-band image is about 0.1 M$_{\odot}$ for $A_v=6$ and about 0.3 M$_{\odot}$ for $A_v=13$.
Similarly, the highest-mass star present in our image is estimated to be between 4 and 7 M$_{\odot}$. This value is in good agreement with the absence of massive stars as inferred from the non-detection of water masers (\cite{codella95}) and the lack of VLA continuum emission. 

In fact, according to the VLA 6~cm observations (section 2.3), and down to the sensitivity of these observations, no radio continuum emission was detected at this wavelength. Setting the lower limit detection for the 6~cm continuum emission at 4$\sigma$,
and adopting a distance of 7~kpc, the rate of ionizing photons required to maintain the ionization of a possible HII region up to that level is $\dot{N_i} \simeq ~3 \times 10^{45}$~s$^{-1}$ (under the assumption of an optically thin, homogeneous spherical HII region with constant temperature of $T=10^{4}$ K; e.g. Rodr\'iguez et al. 1980). This implies an upper limit of $\sim$ 5.5~10$^3$~$L_\odot$ (\cite{panagia73}; corresponding to a B1 zero-age main-sequence star with M~$\sim$~12~M$_{\odot}$) for the luminosity of the individual members of the IRAS 06361-0142 cluster.

Furthermore, the luminosity from the cluster region is dominated by the mid and far-infrared flux as measured by IRAS. We estimate this $L_{\rm FIR}$ to be about $2\times 10^3 L_{\odot}$. Even assuming (unrealistically) that all this luminosity is produced by a single star, this would set an upper limit of about 7 M$_{\odot}$ for any massive star present in this cluster, again in good agreement with the above estimates.

The presence and properties of this young stellar cluster at this location in the far outer Galactic disc corroborates the conclusion of \cite{snell02}, namely that the properties of the far outer Galaxy molecular clouds and their star formation activity are quite similar to those of molecular clouds in the inner Galaxy.

\section{Summary}

We have discovered a new young stellar cluster seen towards a molecular cloud. 
The cloud and clusters are located at a distance of 7 kpc and a 
Galactocentric distance of 15 kpc, well in the far outer Galaxy.
The cluster is detected in our near-infrared images as 
a group of several tens of stars, surrounded by nebulosity of circular shape
and contained in a region of about 1.2 $\times$ 1.2 pc$^2$ centred close to IRAS 06361--0142. It appears composed of low and intermediate-mass young reddened stars with a large fraction having cleared the inner regions of their circumstellar discs. The observations are compatible with a 4 Myr cluster with variable spatial extinction between $A_v=6$ and $A_v=13$. 

These findings indicate that the distant outer Galaxy continues to be active in the production of new stellar clusters, 
with the physical conditions required for the formation of clusters continuing to be met in the very distant environment of the outer Galactic disc.
This suggests that the star formation
process within a molecular cloud is unaffected by the properties that
distinguish the outer Galaxy (e.g. metallicity, surface density) from the solar neighbourhood or the inner Galaxy.

\begin{acknowledgements}
JLY thanks the staff of the Department of Astronomy and Meteorology of the University of Barcelona for hosting him during his sabbatical leave.
This work has been partly supported by the Portuguese Funda\c{c}\~ao
para a Ci\^encia e Tecnologia (FCT). JLY acknowledges partial financial support from the grant AYA2004-05395 from MEC (Spain).
ALS acknowledges financial support from the Ministerio de Educaci\'on y Ciencia (Spain). JMT acknowledges partial financial support from the Spanish grant AYA2005-08523-C03.
This publication makes use of data products from the Two Micron All
    Sky Survey, which is a joint project of the University of Massachusetts
    and the Infrared Processing and Analysis Center/California Institute of
    Technology, funded by the National Aeronautics and Space Administration
    and the National Science Foundation.
\end{acknowledgements}


\begin{thebibliography}{}

\bibitem[Adams, Lada, \& Shu 1987]{adams87} Adams, F. C., Lada, C. J., \& Shu, F. H. 1987, \apj, 312, 788
\bibitem[Clemens 1985]{clemens85} Clemens, D. P.  1985, \apj, 295, 242
\bibitem[Clemens, Sanders, \& Scoville 1988]{clemens88}  Clemens, D.P., 
Sanders, D.B., \& Scoville, N.Z. 1988 , \apj, 327, 139
\bibitem[Codella et al. 1995]{codella95} Codella, C., Palumbo, G. G. C., Pareschi, G., Scappini, F., Caselli, P., \& Attolini, M. R. 1995, \mnras, 276, 57
\bibitem[Digel et al. 1996]{digel96} Digel, S. W., Lyder, D. A., Philbrick, A. J., Puche, D., \& Thaddeus, P. 1996, ApJ, 458, 561
\bibitem[Haisch, Lada, \& Lada 2001]{haisch01} Haisch, K. E., Jr., Lada, E. A., \& Lada, C. J. 2001, \apj,553, 153
\bibitem[Heyer et al. 1998]{heyer98} Heyer, M. H., Snell, R. L., Brunt, C., Howe, J., Schloerb, F. P., \& Carpenter, J. M. 1998, \apjs, 115, 241 
\bibitem[Kobayashi \& Tokunaga (2000)]{kobayashi00} Kobayashi, N. \& Tokunaga, 
A. T.  2000, \apj, 532, 423
\bibitem[Horner, Lada, \& Lada 1997]{horner97} Horner, D. J., Lada, E. A., \& Lada, C. J. 1993, \aj, 113, 1788
\bibitem[Luhman et al. 1998]{luhman98} Luhman, K. L., Rieke, G. H., Lada, C. 
J., \& Lada, E. A. 1998, \apj, 508, 347
\bibitem[McCaughrean \& Stauffer 1994]{mccaughrean94} McCaughrean, M. J., \& 
Stauffer, J. R. 1994, \aj, 108, 1382
\bibitem[Moorwood et al. 1998]{moorwood98}  Moorwood et al. 1998, Messenger, 94, 7
\bibitem[Nota \& Carlson 2007]{nota07} Nota, A. \& Carlson, L. 2007, AAS, in press
\bibitem[Panagia 1973]{panagia73} Panagia, N. 1973, \aj, 78, 929
\bibitem[Rieke \& Lebofsky 1985]{rieke85} Rieke, G. H., \& Lebofsky, M. J. 1985, \apj, 288, 618
\bibitem[Rodr\'{\i}guez et al. 1980]{rodriguez80} Rodr\'{\i}guez, L. F., Moran, J. M., Ho, P. T. P. \& Gottlieb, E. W. 1980, \apj, 235, 845
\bibitem[Santos et al. (2000)]{santos00} Santos, C. A., Yun, J. L., Clemens, D. P., \& Agostinho, R. J. 2000, \apj, 540, 87
\bibitem[Siess, Dufour \& Forestini (2000)]{siess00}  Siess L., Dufour E., \& Forestini M. 2000, \aap, 358, 59
\bibitem[Snell, Carpenter, \& Heyer (2002)]{snell02} Snell, R. L., Carpenter, J. M., \& Heyer, M. H. 2002, \apj, 578, 229
\bibitem[Strom, Strom, \& Merrill 1993]{strom93} Strom, K. M., Strom, S. E., 
\& Merril, M. 1993, \apj, 412, 233
\bibitem[Tapia et al. 1991]{tapia91} Tapia, M., Lop\'ez, J. A., Rubio, M., 
Persi, P., \& Ferrari-Toniolo, M. 1991, \aap, 242, 388
\bibitem[Wood \& Churchwell 1989]{wood89} Wood, D. O. S., \& Churchwell, E. 1989, \apj, 340, 265 
\bibitem[Ybarra \& Phelps 2004]{ybarra04} Ybarra, J. E., \& Phelps, R. L. 2004, \aj, 127, 3444
\bibitem[Yun \& Clemens 1994]{yun94} Yun, J. L., \& Clemens, D. P. 1994, \aj, 
108, 612

\end{thebibliography}
\end{document}